%% file: main.tex
\documentclass[sigconf]{acmart}

\AtBeginDocument{%
  \providecommand\BibTeX{{%
    \normalfont B\kern-0.5em{\scshape i\kern-0.25em b}\kern-0.8em\TeX}}}

\copyrightyear{2024}
\acmYear{2024}
\setcopyright{rightsretained}
\acmConference[ICSE-SEIS'24]{Software Engineering in Society}{April 14--20, 2024}{Lisbon, Portugal}
\acmBooktitle{Software Engineering in Society (ICSE-SEIS'24), April 14--20, 2024, Lisbon, Portugal}\acmDOI{10.1145/3639475.3640110}
\acmISBN{979-8-4007-0499-4/24/04}

\input{preamble.tex}

\begin{document}

\title{Energy Patterns for Web: An Exploratory Study}

\author{Pooja Rani}
\orcid{0000-0001-5127-4042}
\email{rani@ifi.uzh.ch}
\affiliation{%
  \institution{University of Zurich}
  \city{Zurich}
  \country{Switzerland}
}

\author{Jonas Zellweger}
\orcid{0009-0008-5426-4972}
\email{jonas.zellweger@uzh.ch}
\affiliation{%
  \institution{University of Zurich}
  \city{Zurich}
  \country{Switzerland}
}

\author{Veronika Kousadianos}
\orcid{0009-0004-3055-4343}
\email{veronika.wu@students.unibe.ch}
\affiliation{%
 \institution{University of Bern}
 \city{Bern}
\country{Switzerland}
 }
 
 \author{Luis Cruz}
 \email{L.Cruz@tudelft.nl}
\affiliation{%
 \institution{Delft University of Technology }
  \city{Delft}
\country{The Netherlands}
}

\author{Timo Kehrer}
 \email{timo.kehrer@unibe.ch}
\affiliation{%
 \institution{University of Bern}
 \city{Bern, Switzerland}
\country{Switzerland}
 }
 
\author{Alberto Bacchelli}
\email{bacchelli@ifi.uzh.ch}
\affiliation{%
  \institution{University of Zurich, Switzerland}
  \city{Zurich}
  \country{Switzerland}
}
\renewcommand{\shortauthors}{Rani et al.}


\begin{abstract}
As the energy footprint generated by software is increasing at an alarming rate, understanding how to develop energy-efficient applications has become a necessity.
Previous work has introduced catalogs of coding practices, also known as energy patterns.
These patterns are yet limited to Mobile or third-party libraries.
In this study, we focus on the Web domain---a main source of energy consumption.
First we investigated whether and how Mobile \eps can be ported to this domain and found that 20 patterns could be ported.
Then, we interviewed six expert web developers from different companies to challenge the ported patterns.
Most developers expressed concerns for \antipatterns, specifically with functional \antipatterns, and were able to formulate guidelines to locate these patterns in the source code. 
Finally, to quantify the effect of Web \eps on energy consumption, we set up an automated pipeline to evaluate two ported patterns: \ \singlequote{\dynretdel} (\drr) and  \singlequote{\openwneccesary} (\oown). With this, we found no evidence that the \drr pattern consumes less energy than its antipattern, while the opposite is true for \oown. 
Data and Material: \url{https://doi.org/10.5281/zenodo.8404487}
\end{abstract}

\maketitle

\input{content}

\bibliographystyle{ACM-Reference-Format}
\bibliography{references}

\end{document}

%% file: preamble.tex
 \usepackage{amsmath,amsfonts}
 \usepackage{textcomp}
\usepackage{xspace}
\usepackage[normalem]{ulem} 
\usepackage{url} 

\usepackage{subcaption}
\usepackage{tablefootnote}
\usepackage{array, booktabs}
\usepackage{blindtext}
\usepackage{ifthen}
\usepackage[shortlabels]{enumitem}

\usepackage{csvsimple}
\usepackage{supertabular} 
\usepackage[most]{tcolorbox}

\usepackage{tabularx}

\usepackage{fontawesome}      
\usepackage{adjustbox} 		
\usepackage{pifont} 		

\newcolumntype{R}[2]{
    >{\adjustbox{angle=#1,lap=\width-(#2),margin*=0.4em 0em 0em 0em}\bgroup}
    l
    <{\egroup}
}

\let\oldFootnote\footnote
\newcommand\nextToken\relax
\renewcommand\footnote[1]{%
    \oldFootnote{#1}\futurelet\nextToken\isFootnote}
\newcommand\isFootnote{%
    \ifx\footnote\nextToken\textsuperscript{,}\fi}

\newcommand{\checkdecimal}[1]{%
    \IfDecimal{#1}{#1}{0}%
}

\definecolor{gray50}{gray}{.5}
\definecolor{gray40}{gray}{.6}
\definecolor{gray30}{gray}{.7}
\definecolor{gray20}{gray}{.8}
\definecolor{gray10}{gray}{.9}
\definecolor{gray05}{gray}{.95}

\newlength\Linewidth
\def\findlength{\setlength\Linewidth\linewidth
	\addtolength\Linewidth{-4\fboxrule}
	\addtolength\Linewidth{-3\fboxsep}
}
\newenvironment{rqbox}{\par\begingroup
	\setlength{\fboxsep}{5pt}\findlength
	\setbox0=\vbox\bgroup\noindent
	\hsize=0.95\linewidth
	\begin{minipage}{0.95\linewidth}\normalsize}
	{\end{minipage}\egroup
	\textcolor{gray20}{\fboxsep1.5pt\fbox
		{\fboxsep5pt\colorbox{gray05}{\normalcolor\box0}}}
	\endgroup\par\noindent
	\normalcolor\ignorespacesafterend}

\newcounter{Finding}
\stepcounter{Finding}

\newcommand{\ie}{\emph{i.e.},\xspace}
\newcommand{\eg}{\emph{e.g.},\xspace}
\newcommand{\etal}{\emph{et al.}\xspace}
\newcommand{\etc}{\emph{etc.}\xspace}
\newcommand{\github}{{GitHub}\xspace}
\newcommand\singlequote[1]{`{#1}'\xspace}

\newcommand{\darkui}{Dark UI Colors\xspace}
\newcommand{\dynretdel}{Dynamic Retry Delay\xspace}
\newcommand{\avoidextrawork}{Avoid Extraneous Work\xspace}
\newcommand{\raceidle}{Race-to-idle\xspace}
\newcommand{\openwneccesary}{Open Only When Necessary\xspace}
\newcommand{\pushpoll}{Push over Poll\xspace}
\newcommand{\powersave}{Power Save Mode\xspace}
\newcommand{\poweraware}{Power Awareness\xspace}
\newcommand{\reducesize}{Reduce Size\xspace}
\newcommand{\wificell}{WiFi over Cellular\xspace}
\newcommand{\suplog}{Suppress Logs\xspace}
\newcommand{\batchoper}{Batch Operations\xspace}
\newcommand{\cache}{Cache\xspace}
\newcommand{\decreaserate}{Decrease Rate\xspace}
\newcommand{\usrknowbest}{User Knows Best\xspace}
\newcommand{\informusr}{Inform Users\xspace}
\newcommand{\enoresolution}{Enough Resolution\xspace}
\newcommand{\sensorfusion}{Sensor Fusion\xspace}
\newcommand{\killtask}{Kill Abnormal Tasks\xspace}
\newcommand{\noscrinteraction}{No Screen Interaction\xspace}
\newcommand{\avoidgraphic}{Avoid Extraneous Graphics and Animations\xspace}
\newcommand{\mansyncod}{Manual Sync – On Demand\xspace}

\newcommand{\eps}{energy patterns\xspace}
\newcommand{\eaps}{energy antipatterns\xspace}
\newcommand{\EPS}{Energy Patterns\xspace}

\newcommand{\antipattern}{antipattern\xspace}
\newcommand{\antipatterns}{antipatterns\xspace}

\newcommand{\powergadget}{Intel Power Gadget\xspace}
\newcommand{\rapl}{Running Average Power Limit\xspace}
\newcommand{\drr}{DRD\xspace}
\newcommand{\oown}{OOWN\xspace}

\newcommand{\yes}{\ding{52}}
\newcommand{\no}{\ding{54}}

\newcolumntype{L}[1]{>{\raggedright\arraybackslash}m{#1}} 

\def\BibTeX{{\rm B\kern-.05em{\sc i\kern-.025em b}\kern-.08em
    T\kern-.1667em\lower.7ex\hbox{E}\kern-.125emX}}
    
\newboolean{showedits}
\setboolean{showedits}{true} 
\ifthenelse{\boolean{showedits}}
{
	\newcommand{\del}[1]{\textcolor{red}{\sout{#1}}} 
	\newcommand{\nbe}[3]{
		{\colorbox{#3}{\bfseries\sffamily\scriptsize\textcolor{white}{#1}}}
		{\textcolor{#3}{\sf\small$\blacktriangleright$\textit{#2}$\blacktriangleleft$}}}
}{
	\newcommand{\del}[1]{} 
	
	\newcommand{\nbe}[3]{}
}

\ifthenelse{\boolean{showedits}}
{
	\newtcolorbox{inserted}{
		title=Inserted text:,
		colframe=blue,colback=blue!5!white,
		breakable,
		leftrule=0mm, 
		bottomrule=0mm,
		rightrule=0mm,
		toprule=0mm,
		arc=0mm, outer arc=0mm,
		oversize
	}
	\newtcolorbox{deleted}{
		title=Deleted text:,
		colframe=red,colback=red!5!white,
		breakable,
		leftrule=0mm, 
		bottomrule=0mm,
		rightrule=0mm,
		toprule=0mm,
		arc=0mm, outer arc=0mm,
		oversize
	}
	\newtcolorbox{refactored}{
		title=Rewritten text:,
		colframe=blue,colback=red!5!white,
		breakable,
		leftrule=0mm, 
		bottomrule=0mm,
		rightrule=0mm,
		toprule=0mm,
		arc=0mm, outer arc=0mm,
		oversize
	}
}{

}

\newboolean{showcomments}
\setboolean{showcomments}{true}
\ifthenelse{\boolean{showcomments}}
{\newcommand{\nbc}[3]{
		{\colorbox{#3}{\bfseries\sffamily\scriptsize\textcolor{white}{#1}}}
		{\textcolor{#3}{\sf\small$\blacktriangleright$\textit{#2}$\blacktriangleleft$}}}
	}
{\newcommand{\nbc}[3]{}
	}

\definecolor{source}{gray}{0.9}
\newboolean{isblinded}
\setboolean{isblinded}{true}
\ifthenelse{\boolean{isblinded}}
{\newcommand\blind[1]{BLINDED\xspace}}
{\newcommand\blind[1]{#1\xspace}}

\newcommand{\rqI}{To what extent can we port energy patterns for mobile applications to web applications?}
\newcommand{\rqII}{How do developers perceive the ported web \eps and locate them in source code?}
\newcommand{\rqIII}{How do \emph{\dynretdel} and \emph{\openwneccesary} impact energy consumption?}


%% file: content.tex
\section*{Lay Abstract}
The information technology sector significantly affects the climate. 
With our increasing online activities, from chatting to accessing medical history, software powering these services requires to be energy-efficient. 
Researchers in software engineering have been exploring green coding practices, or energy-specific design patterns (aka \eps) to make software more eco-friendly. 
While such energy practices have been explored for other domains including Mobile, Web applications have been somewhat overlooked, despite our daily heavy internet use.
We focused on the existing \eps from Mobile applications to Web applications.  
To validate these ported \eps, we interviewed six professional web developers from various companies. 
Then, we tested some patterns to see if these energy patterns indeed save any energy. Our results showed that developers are unaware of the energy practices and some patterns did not make a noticeable difference, while others consume more energy than their counterpart.
In a nutshell, our work highlights the knowledge gap between green coding research and industry and emphasize the need to understand the trade-offs in energy practices for  sustainable digital future.

\section{Introduction}
The ICT sector is estimated to generate up to 5.5\% of world carbon emissions and to consume 20\% of all electricity~\cite{andrae2015global}. 
Indeed, from healthcare to communication, every industry prominently runs on software, thus understanding and developing energy-efficient software is urgent.

In this context, the Software Engineering (SE) research community has started investigating \emph{green coding} and \eps for source code~\cite{manotas2016empirical,georgiou2019software}.
Energy-specific design patterns for source code (henceforth, \emph{\EPS}) are best practices developers use to make their source code energy-efficient~\cite{manotas2016empirical}. While researchers have developed catalogs of \eps for Mobile applications \cite{cruz2019catalog} and for deep learning libraries \cite{Shanbhag2022}, some domains are still yet to be covered, prominently the Web domain, which is particularly relevant as its energy consumption is ever increasing~\cite{koomey2007estimating}.



Our goal is to gather and evaluate Web-specific \eps. To this aim, we first attempt to port existing Mobile \eps \cite{cruz2019catalog} to the Web domain. 
Then, to challenge our ported patterns, we discuss them with six professional Web developers, by means of in-depth structured interviews.
In particular, we discuss how understandable these patterns are, how they are perceived, and whether they can be located in source code of Web applications. Consequently, we collect concerns regarding various patterns or their respective \antipatterns (not having the pattern) and guidelines for locating them in the source code. Based on the guidelines, we analyze the source code of a company 
and thus show which of them are easy to locate or not.  
Finally, we measure the impact of two \eps to see whether they indeed save energy.

Our results show that most Mobile \eps (16 patterns) can be directly ported to the Web, while a few (four patterns) are only partially applicable to Web applications (\eg \singlequote{\powersave}), and two of them cannot be ported because they are inapplicable (\eg \singlequote{\wificell}).
From the interview study, we start by confirming that past results on developers' limited awareness of energy aspects~\cite{Pang2016, Yamashita2013} also apply to our Web developers.
Once we described the patterns, participants report to understand them and many expressed concerns for \eaps, specifically for functional \antipatterns.
Developers then were able to provide guidelines to locate the patterns in the source code. Following their guidelines, we could identify eleven \antipatterns and nine patterns in their own source code.   
Some \eps, \eg  \singlequote{\pushpoll}, could be easily found in the source code, while others, \eg   \singlequote{\reducesize},  \singlequote{\enoresolution} were harder to identify.

Finally, concerning the energy impact of \singlequote{\dynretdel} and  \singlequote{\openwneccesary}, we found no evidence that the \drr pattern consumes less energy than its antipattern, while the opposite is true for \oown.

\smallskip
With this paper, we make the following main contributions:
\begin{itemize}
    \item a porting of existing \eps for Mobile to the Web domain;
    \item a report on the perception of six professional web developers on the ported \eps;
    \item a guideline to find Web \eps in source code based on developers' experience and confirmed against these developers' code bases;
    \item empirical evidence on the effects of \antipatterns on energy consumption for  \singlequote{\dynretdel} and  \singlequote{\openwneccesary};
    \item a replication package (RP)~\cite{reppackage} containing detailed patterns with context and discussions, interview results, and an automated pipeline to quantitatively evaluate \eps.
\end{itemize}
 

\section{Study Design}

In this section we motivate and present our research questions and detail the research method we use to answer them.

\subsection{Research questions}

While there's been a growing focus on green coding practices in Mobile and embedded systems~\cite{manotas2016empirical,Shanbhag2022}, the Web domain remains under-explored. Cruz \etal developed a catalog of 22 Mobile \eps 
\cite{cruz2019catalog}.
Our aim is to expand current scientific understanding by exploring the portability of these Mobile \eps to Web applications. Insights from this study can equip Web developers with tools to integrate \eps into their codebase and recognize \antipatterns (inefficient practices) to minimize energy use. Hence we ask:

\begin{center}
	\begin{rqbox}
		\begin{description}
			\item $RQ_1$: \rqI
		\end{description}
	\end{rqbox}
\end{center}

Developers are the main target of \eps, because these patterns---akin to design patterns---should provide a roadmap for developers to code refactoring and reduced energy consumption. So far, previous work that has proposed catalogs of \eps \cite{cruz2019catalog,Shanbhag2022, albonico2021mining}, did not investigate how professional developers perceive and understand these patterns, and how easily they can be located in source code.
We try to fill this gap in our context: We actively engage industrial developers, inviting them to assess and challenge the \eps we have ported from the Mobile domain to the Web one. Through this endeavor, we aim to shed light on their perspectives and hurdles regarding energy consumption in software, the ported \eps, as well as to measure their commitment to dedicating resources for improving the energy efficiency of Web applications. We ask:

\begin{center}
	\begin{rqbox}
		\begin{description}
			 \item  $RQ_2$: \rqII
		\end{description}
	\end{rqbox}
\end{center}

In \texorpdfstring{$RQ_1$}, we ported Mobile \eps to Web applications. 
Whether these \eps indeed impact the energy consumption of a Website is not explored.
Previous work has investigated the energy impact of software specific to its code ~\cite{carccao2014measuring}, execution \cite{pinto2014understanding,ribic2014energy}, third-party libraries \cite{schuler2020characterizing}, programming languages \cite{pereira2017energy}, and Mobile applications \cite{kwon2013reducing}.
Researchers have also measured the impact of test smells or code smells and their associations with energy consumption.
We aim to investigate the actual impact of our ported \eps, as validated by the professional developers.
Such analysis can help practitioners make informed choices to reduce the energy impact of their Web applications. We focus on two specific patterns (for the reasons mentioned later) on and ask:

\begin{center}
	\begin{rqbox}
		\begin{description}
			 \item $RQ_3$: \rqIII
		\end{description}
	\end{rqbox}
\end{center}

\subsection{Methodology}
\subsubsection{\texorpdfstring{$RQ_1$}{RQ1}: \EPS for The Web} 
Previous work has identified various coding practices for maintaining source code (\eg design patterns), or improving the energy consumption (\eg \eps).
\citet{cruz2019catalog} have identified a set of \eps developers apply intending to improve the energy aspect of their applications.  They have analyzed 1,783 Mobile applications from both Android and iOS
from \github. 
They identified 22 energy patterns appearing in 421 instances \cite{cruz2019catalog} (described in the RP \cite{reppackage}).
Since they provide a recent and the largest catalog of such patterns, we attempt to port these patterns to the Web.
We 
investigated if and how many of them are applicable on the Web domain.
To port these energy patterns to the Web, we restricted our focus to a client-server architecture
for the Web since various Web architectures or frameworks can have various implementations of the patterns. 

For each pattern, one author studied the original pattern from \citet{cruz2019catalog} with its description and examples and attempted to formulate the Web counterpart.
To find the Web example, she 
checked the code examples given by Cruz \etal in their RP and 
searched similar code snippets or the pattern name if exists
into the Google search engine. 
Since the goal is to find similar concepts or examples, she also looked at gray literature like guidelines, documentation, and blogs.
Once the patterns were formulated, another author reviewed each definition with its examples and pointed out the differences. We found the differences mainly in definitions of 10 patterns, \eg `\darkui' in Mobile focuses on saving battery for AMOLED screens while for Web, battery is not the primary focus, therefore we phrased the Web pattern accordingly. The detailed discussion on differences is provided in Appendix in the RP. The third author reviewed the differences and resolved them by mutual discussion. This process led to \eps for the Web, with description and examples.
Certain patterns exist in the front-end part of the application, while others exist in the back end; we dissected them based on their existence in the front- or back-end part of the application. This helps identify which patterns require access to source code and which can be identified by merely exploring the Website. 

\subsubsection{\texorpdfstring{$RQ_2$}{RQ2}: Developers' Perceptions on Web \EPS}

 \begin{figure*}[ht]
	\includegraphics[width=\linewidth]{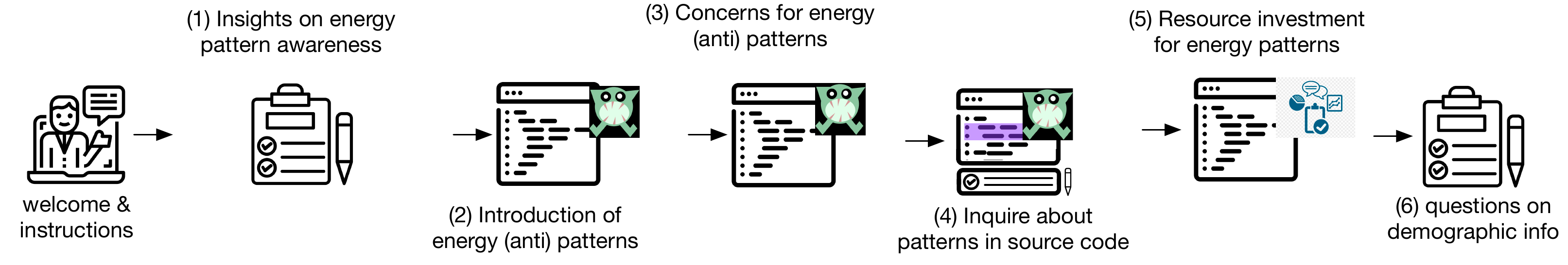}
	\caption{Interview study design to answer RQ$_2$: Industry Insights on \EPS}
	\label{fig:rq2-study-flow}	
\end{figure*}

To challenge our mapped patterns in practice, we investigated the perception and concerns of professional developers.
Following a convenience sampling approach, we interviewed six expert Web developers from four different companies. 
Three of them work at the same company,\footnote{The identity of the company cannot be disclosed due to a non-disclosure agreement.} thus allowing us to perform a case study.
The company provides a web portal where the customers can see their electricity consumption data. The portal is a web application with a client-server architecture developed mainly using the front end technologies: \texttt{NodeJs}, \texttt{VueJs}, \texttt{Element Plus}, \texttt{TypeScript}, \texttt{Axios}, and the back end technologies: \texttt{Python} and \texttt{Django} framework. The whole web application consists of 57,000 lines of code (LoC) for the front end and around 1,590,000 LoC for the back end. The three other developers we interviewed belong to three different companies yet focus on similar technologies.

Various methods have been used to investigate developers' perception of energy aspects, such as mining software ecosystems or surveying developers \cite{georgiou2019software,Yamashita2013}. 
\citet{fink2003survey} explained that ``\emph{the interview enables to collect information from people to explain their attitude, behavior, and knowledge}.'' Therefore we considered it as the best approach to collect our data.
\autoref{fig:rq2-study-flow} shows the design of our interview study, as inspired by \citet{fink2003survey} and \citet{Yamashita2013}.
We followed various steps to set up the interview: (a) establish the goals and questions, (b) prepare an interview instrument, (c) organize the interview administration, and (d) analyze the responses and collect the guidelines based on them. 

\paragraph{Establishing the Goals and Questions}
Three authors outlined the goals, mainly focusing on challenging the ported \eps with the help of professional developers. We also aimed at collecting guidelines to identify the valid patterns in their source code, and at understanding whether they would invest resources to implement \eps or remove \antipatterns.
Therefore, we defined intermediate goals, \eg (1) insights on energy pattern awareness, (2) introduction of patterns as shown in \autoref{fig:rq2-study-flow}. 
%
Based on the main goals, we chose an \emph{exploratory structured interview} that included both closed and open questions to understand their motivation or choices like why certain \eps concerned them.

\paragraph{Preparing Interview Instrument}
We began by asking about their awareness of \eps and their sources of information. 
Then, we challenged the ported patterns by introducing them to developers, examples to clarify, and their \antipatterns. 
We gauged their level of concern regarding the presence of \antipatterns.
We used a five-point Likert scale and designed various open-ended questions to understand their reasons of concerns, where in the source code these \eps can be found, how easy it is to identify them, 
their desired commitment to invest resources, and their past use of related tools.
At the end, we asked demographic questions to understand their background, \eg role in the organization, years of experience in Python, JavaScript, and TypeScript (languages used in the company) 
and the programming paradigms as certain patterns can be specific to a language specifically in trusting the guidelines proposed by developers to search \eps in source code. The
questions are provided in RP~\cite{reppackage}.

\paragraph{Organizing the Interview}
We conducted a personally administered interview using anonymous Google Forms. 
The questions were asked in person and developers filled their responses directly.
For the longer open-ended questions (\eg where these \eps can be found in their source code), the author transcribed developer responses on-site to avoid recording the sessions, which would have raised confidentiality concerns.

\paragraph{Analyzing the Responses and Guidelines}
We visually analyzed the graphs of the closed-ended questions, while, for the open-ended ones, we conducted axial and open coding \cite{strauss1998basics}.
That is, we extracted themes and excerpts from the developer's responses. 
Since one question focused on collecting the guidelines, each author independently analyzed half of the responses and reviewed the other half. 
Both authors handled their disagreement by mutual discussion.
Finally,  we verified the guidelines by checking the presence or absence of \eps in the source code of the company.

\subsubsection{\texorpdfstring{$RQ_3$}{RQ3}: Impact of \EPS} 
 \begin{figure*}[ht]
	\includegraphics[width=\linewidth]{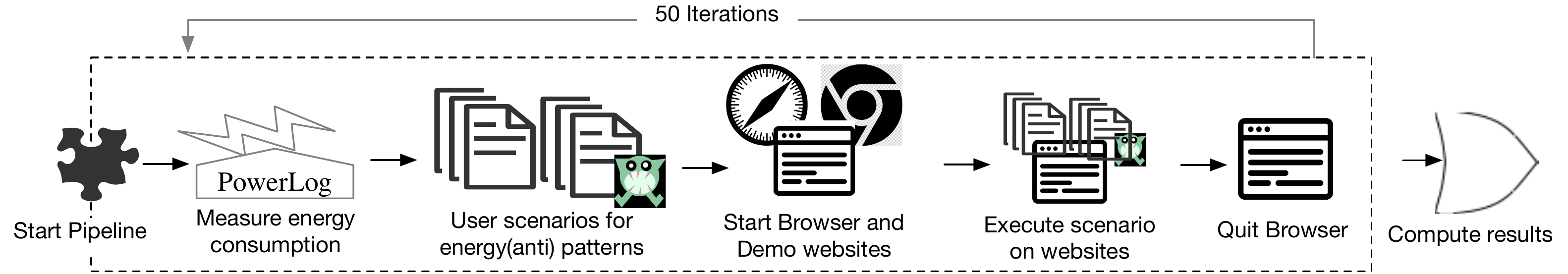}
	\caption{Study design to answer RQ$_3$: Impact of \EPS}
	\label{fig:rq3-study-flow}	
\end{figure*}
Here, we investigated whether \eps indeed impact the energy consumption of a Website.
\citet{ghaleb2019software} showed the accuracy of software tools in estimating energy consumption.
Therefore, to estimate power consumption, we used Intel PowerLog version 3.7.0 which relies on \rapl (RAPL) interface. This approach, widely used in green software studies~\cite{bruce2015reducing,carccao2014measuring,georgiou2019software}, measures real-time power information of macOS based on the energy counters in the Intel Core processor without any hardware instrumentation. 
However, it does not measure energy use of network traffic, display, or an isolated process. 
Consequently, certain patterns could not be measured. Various criteria are detailed below to include or exclude patterns.

We included each pattern that:
  \setlist{nolistsep}
\begin{enumerate}[label=I\arabic*:, start=1, itemsep=0em]
    \item is processor-dependent 
    \item is measurable on the client-side browser
\end{enumerate}

We excluded each pattern that:
\begin{enumerate}[label=E\arabic*:, start=1, noitemsep]
    \item is Mobile device-specific (\ie  \singlequote{\powersave},  \singlequote{\poweraware},  \singlequote{\raceidle},  \singlequote{\wificell},  \singlequote{\sensorfusion},  \singlequote{\noscrinteraction}),
    \item is network related as the selected energy profiler does not measure network impact (\ie  \singlequote{\reducesize},  \singlequote{\batchoper},  \singlequote{\cache},  \singlequote{\pushpoll},  \singlequote{\enoresolution}),
    \item is display-specific (\ie  \singlequote{\noscrinteraction},  \singlequote{\darkui}, or  \singlequote{\avoidgraphic}),
    \item is user-related as it can not be controlled which parameters to change (\ie  \singlequote{\usrknowbest},  \singlequote{\informusr},  \singlequote{\mansyncod}),
    \item is power model related as it affects the whole device and thus is not controllable for only browsers (\ie  \singlequote{\powersave},  \singlequote{\poweraware}),
     \item can not be simulated systematically or lacks empirical data to support its parameter selection (\eg what threshold of logs should be considered for the \suplog pattern, what amount of notification should be considered for  \singlequote{\pushpoll}, what rate for  \singlequote{\decreaserate}, and which tasks should be considered abnormal for  \singlequote{\killtask)},
    \item similar patterns but no specific Web code could be extrapolated, \eg  \singlequote{\avoidextrawork} and  \singlequote{\avoidgraphic} are similar to  \singlequote{\openwneccesary}, but could not find precise web code for it.
\end{enumerate}

Based on these criteria, which ensured the most reliable measurements with the current knowledge, we were left with two patterns, namely  \singlequote{\dynretdel} (\drr) and  \singlequote{\openwneccesary} ({\oown}).

\paragraph{Pipeline Setup}
We automated our test process as shown in \autoref{fig:rq3-study-flow}, using Python.  
The test process initiates PowerLog with Selenium, simulating a user scenario on the demo website. 
PowerLog measures the energy consumption of the executed commands storing parameters in log files. 
These user scenarios are coded in Java, and the website interaction is automated using Selenium WebDrivers (for Chrome and Safari).
The browsers ran in incognito mode to ensure no cache or past data usage.
The demo Websites built using HTML, JavaScript, and PHP,  enabled us to simulate the scenarios for the selected patterns.
We analyzed the generated logs to obtain the results.

\paragraph{Implementation of User Scenarios}
To measure the impact of each pattern, we simulated two user scenarios, 
as shown in \autoref{fig:rq3-study-flow}.

\begin{description}[leftmargin=0.3cm]
    \item[\textbf{\drr:}]

We open the browser, visit the demo website, initiate an HTTP request to an external Web server,
which waits for two seconds and then returns an HTTP 503 error code (the server is unavailable to handle the request or the attempt to access the resource failed).
Then, the Website retries to access the server.
In the pattern scenario, the retry interval increased exponentially after each failed attempt (referred to as \emph{dynamic retry}), while in \antipattern scenario, it remained constant (static retry).
Dynamic retry interval is based on the \emph{EcoAndroid} tool \footnote{\url{https://plugins.jetbrains.com/plugin/15637-ecoandroid}}, \ie 1 second, 2 seconds, 4 seconds \etc
and the static retry interval is fixed as 3 seconds.
The browser quits after 90 seconds to ensure that the scenarios do not execute infinitely.

\item[\textbf{\oown:}] We focused on content loading (specifically images). 
In the pattern scenario,
the website loads images when the user scrolls or enters the viewpoint (\emph{lazy} loading) and defers loading non-visible content. 
In contrast, the \antipattern scenario loaded all images at once when the user opens the website (\emph{eager} loading).
Both scenarios lasted approximately a similar time, \ie 45--48 seconds, and then
the browser quits. 
\end{description}

\paragraph{Execution of User Scenarios}

We ran the pipeline on a Mac mini desktop computer with a 2.3 GHz Quad-Core Intel\textsuperscript{\textregistered} Core\textsuperscript{\texttrademark} i7 processor, 16 GB DDR3 RAM with 1600 MHz, 1 TB SATA HDD, running macOS Catalina 10.15.7.
We run the scenario on two browsers, \ie Chrome version 113.0.5672.92 and Safari version 15.6.1. 
For each scenario, we collected \texttt{Cumulative energy} (energy consumption of the processor), 
\texttt{elapsed time} and (\texttt{Package Temperature}) in the log files \cite{powergadget}.

To mitigate energy consumption variability, each scenario was run for 50 iterations for both pattern and \antipattern for each browser, 
thus 400 times; this took approximately 14 hours. Overall, we took several preventative measures:
\begin{enumerate}[label=P\arabic*:, start=1, noitemsep]
\item Running each scenario for 50 iterations with a 30-second interval (cool-down periods) 
for temperature consistency.
\item Deactivating all non essential applications,
\eg including WiFi, Bluetooth.
\item Limiting the number of accessories, \eg Mac mini is not integrated with a keyboard, mouse, or display.\footnote{The pipeline was started with a remote connection (Screen sharing option in macOS), and the remote connection was disconnected as soon as the pipeline started.}
\item Disabling sleep mode (set the sleep time to never).
\item Randomizing the scenarios of a pattern to avoid external influences. 
\item Keeping the computer continuously plugged in.

\end{enumerate}



\section{Results}
\subsection{\texorpdfstring{$RQ_1$}{RQ1}: Web \EPS} 
We could port 20 Mobile \eps to the Web domain, as shown in \autoref{tab:patterns_adapatability}. 
Of these, 16 patterns fit web applications directly, four patterns mapped partially, and two patterns did not apply (particularly if the website is accessed via desktop or laptop rather than Mobile).  
Patterns like \singlequote{\sensorfusion} and  \singlequote{\wificell} are Mobile-specific as they rely on 
features such as accelerometer or cellular networks.
As the websites are also accessed on desktop or laptop, we marked these patterns as inapplicable.
The porting of patterns required understanding their applicability and trade-offs in specific scenarios, \ie which patterns should be considered for what applications and when.

\autoref{tab:patterns_adapatability} shows the patterns classified based on whether they are relevant to the client-side (front end) or server-side (back end) of a Web application.
This information also helps pinpoint where these patterns can be found. For example client-side patterns (\eg \singlequote{\darkui}, \singlequote{\powersave}, \singlequote{\usrknowbest}) can be found by looking at the website without inspecting the source code.\smallskip

\input{Listings/avoidExtraWork}
\input{tables/patterns-adaptability}

\autoref{tab:patterns_adapatability} provides a breakdown of each adapted pattern accompanied by a representative example. As an illustration, the Mozilla's Page Visibility API, showcased in \autoref{AvoidExtraenousWorkCode}, exemplifies the \singlequote{\avoidextrawork} pattern. This API informs users if a web page is currently visible, enabling functionalities like pausing audio or video when the page is not in view. Moreover, this API also embodies the \singlequote{\oown} pattern, where resources, like audio, are activated only when necessary—specifically, when the user can see the page. These examples highlight the manifestation of such patterns on the Web, guiding us in spotting them in future web applications.

\subsection{\texorpdfstring{$RQ_2$}{RQ2}: Developers on Web \EPS} 
 \begin{figure}[ht]
	\includegraphics[width=\linewidth]{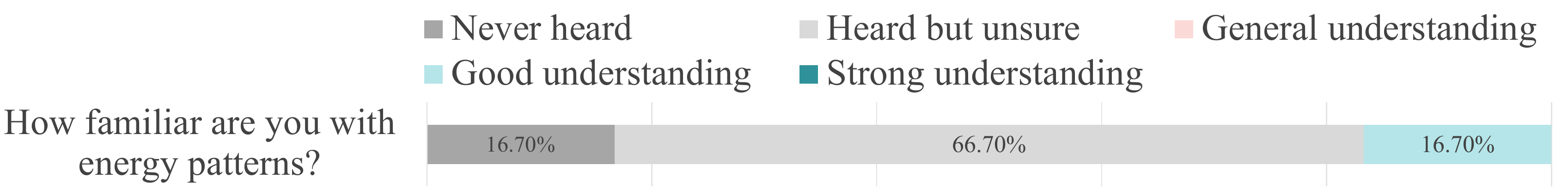}
 	\includegraphics[width=0.8\linewidth]{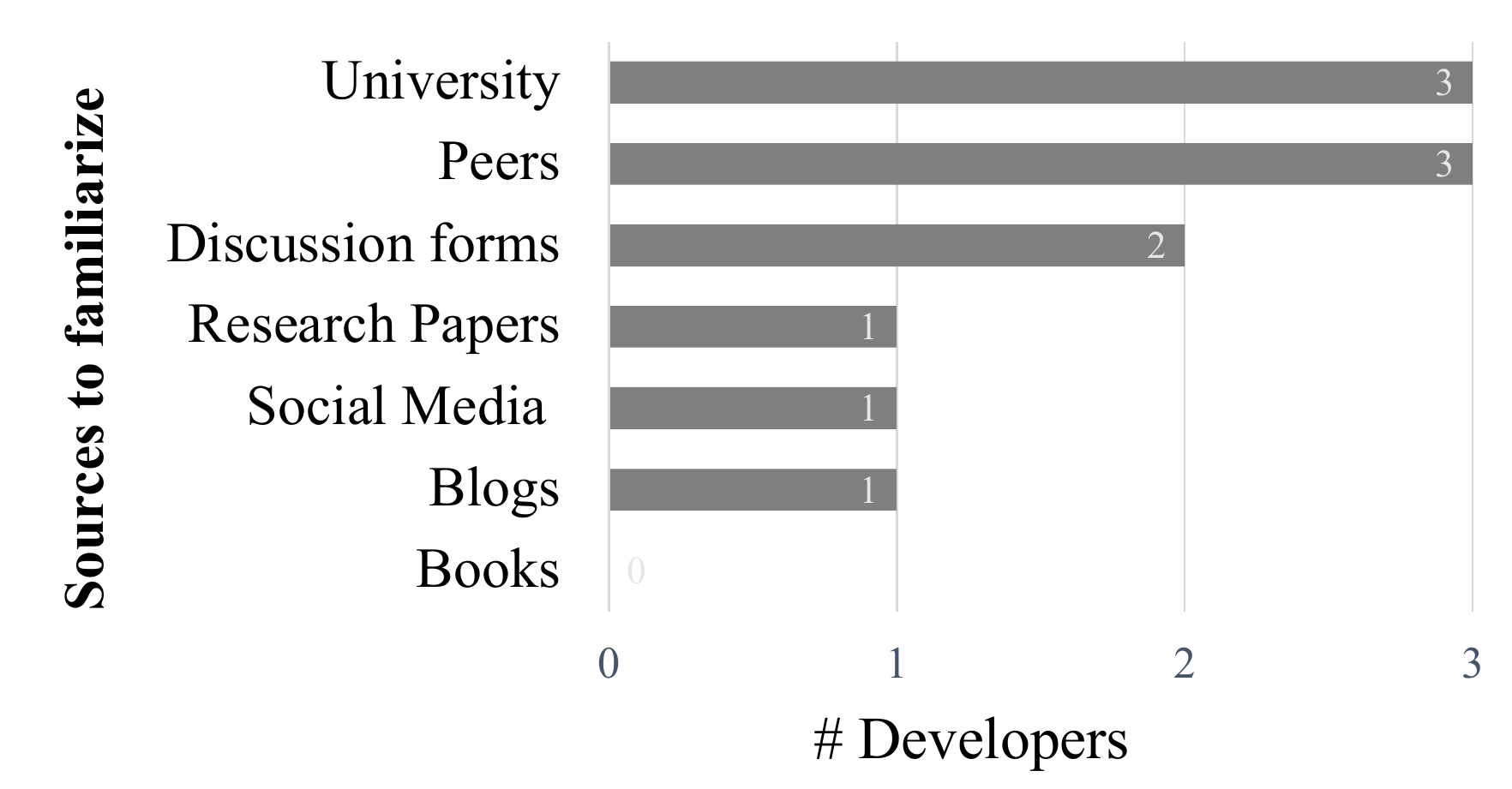}
	\caption{Awareness and sources of Energy Patterns}
	\label{fig:rq2-results-familarity}	
\end{figure}
\begin{figure*}[ht]
     \begin{subfigure}[c]{0.35\linewidth}
     \centering
\includegraphics[width=\linewidth]{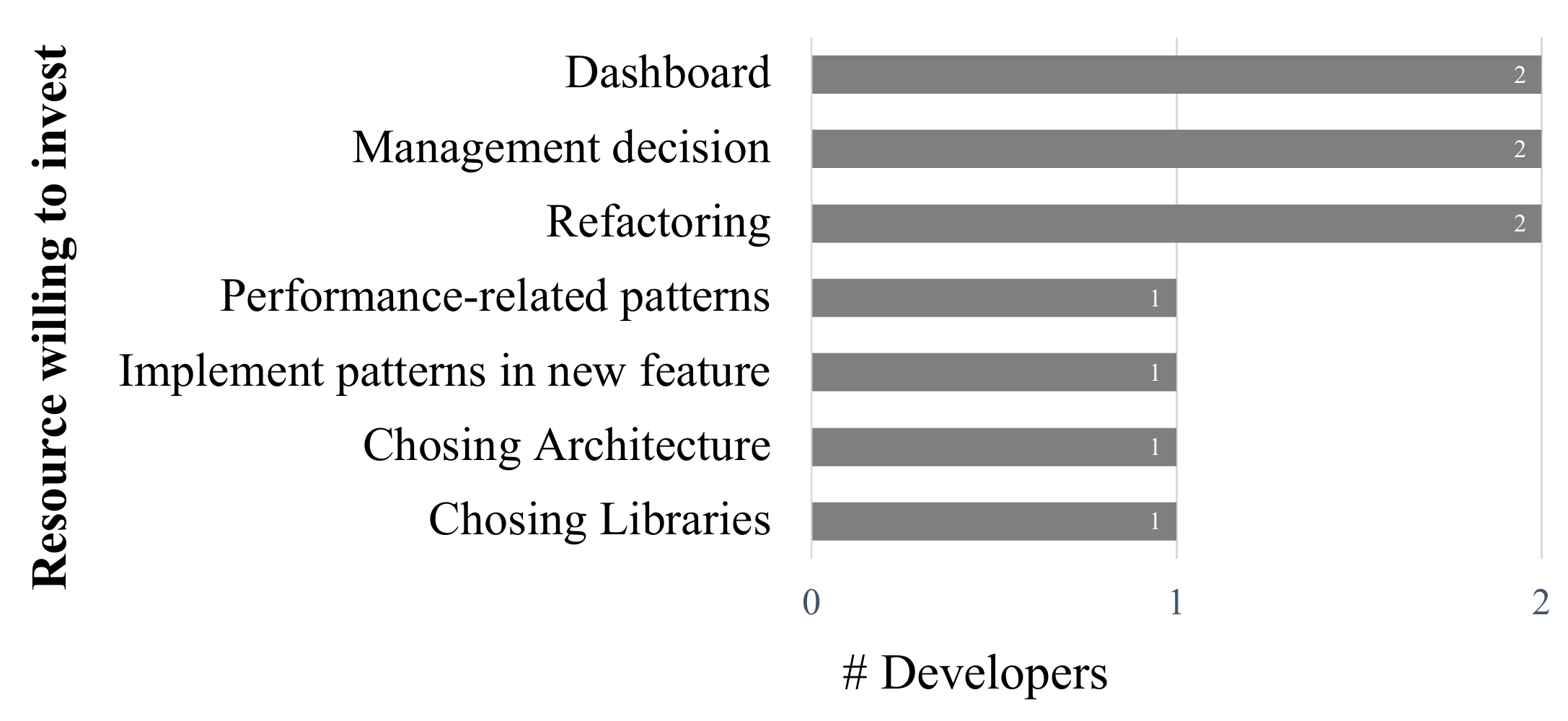} 
	\end{subfigure}\hfill
  \begin{tabular}[c]{@{}c@{}}
	\begin{subfigure}[c]{0.63\linewidth}
 \centering
\includegraphics[width=0.97\linewidth]{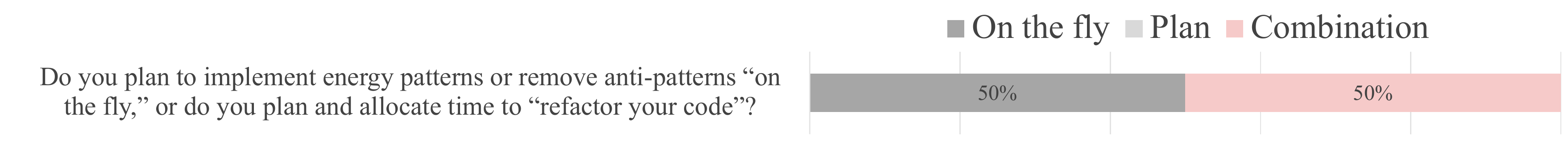} 
	\end{subfigure}%
\\  
\noalign{\bigskip}%
  \begin{subfigure}[c]{0.63\linewidth}
  \centering
  \includegraphics[width=\linewidth]{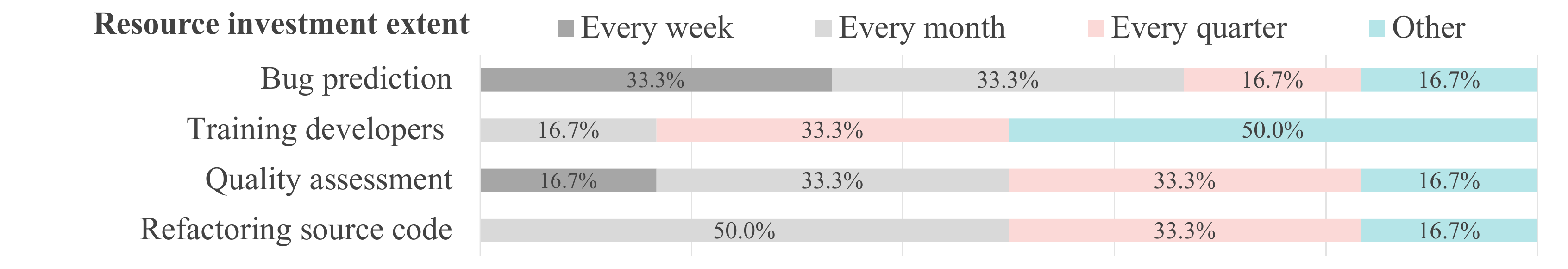}  
	\end{subfigure}%
  \end{tabular}
	\caption{Willingness of developers in investing resource for \eps}
	\label{fig:rq2-results-resource-invest}
\end{figure*}

\begin{figure*}[ht]
     \begin{subfigure}[c]{0.35\linewidth}
     \centering
\includegraphics[width=\linewidth]{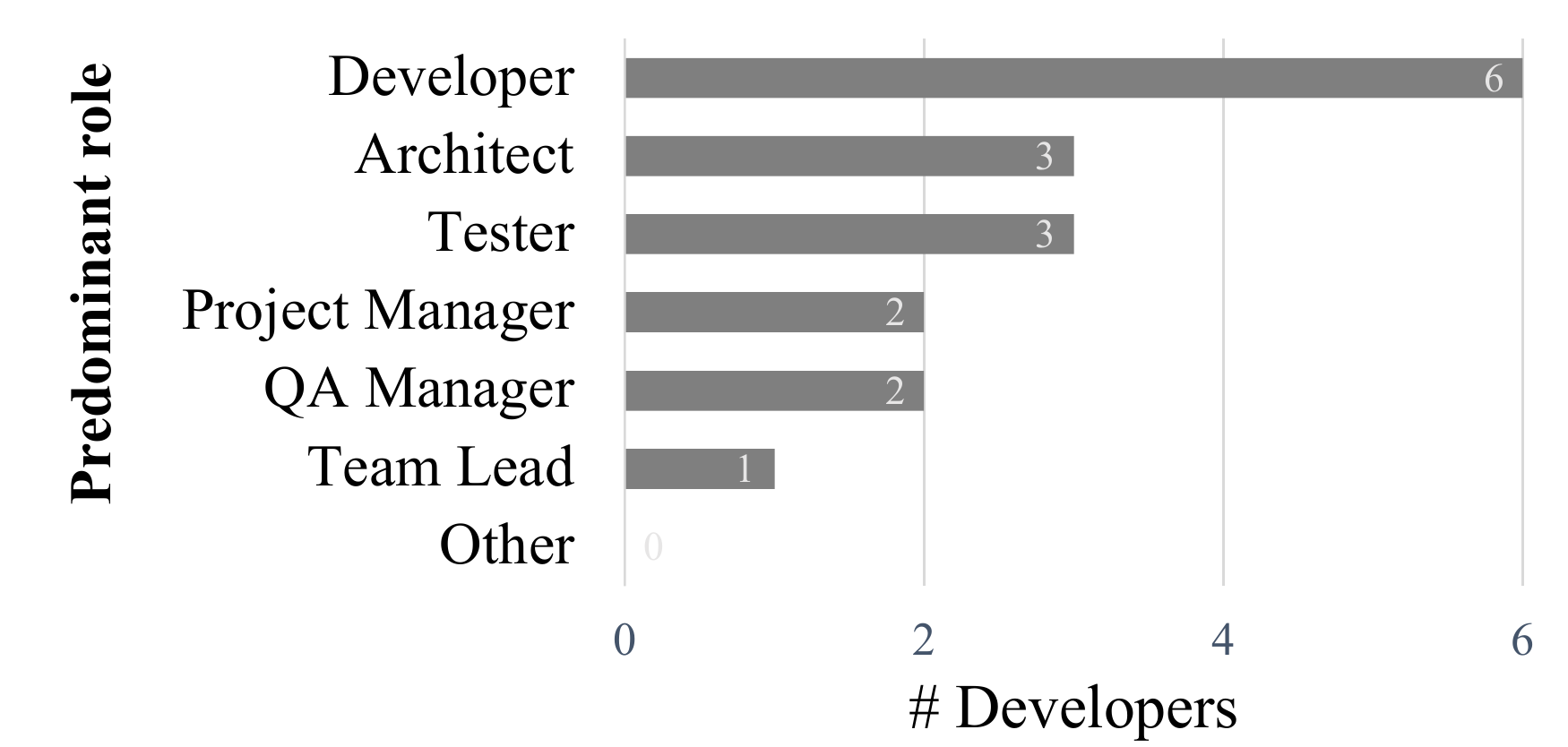} 
	\end{subfigure}\hfill
  \begin{tabular}[c]{@{}c@{}}
	\begin{subfigure}[c]{0.63\linewidth}
 \centering
\includegraphics[width=0.99\linewidth]{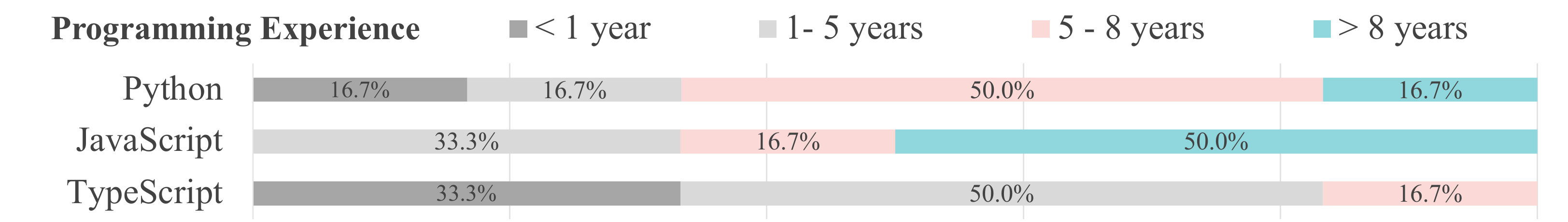} 
	\end{subfigure}%
\\  
\noalign{\bigskip}%
  \begin{subfigure}[c]{0.62\linewidth}
  \centering
  \includegraphics[width=\linewidth]{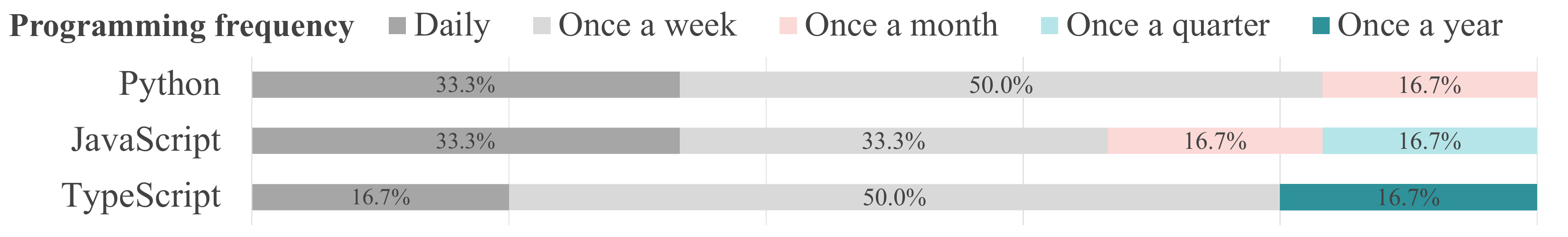}  
	\end{subfigure}%
  \end{tabular}
	\caption{Participants' demographics}
	\label{fig:rq2-results-demographics}
\end{figure*}

\begin{figure}[ht]
	\includegraphics[width=0.98\linewidth]{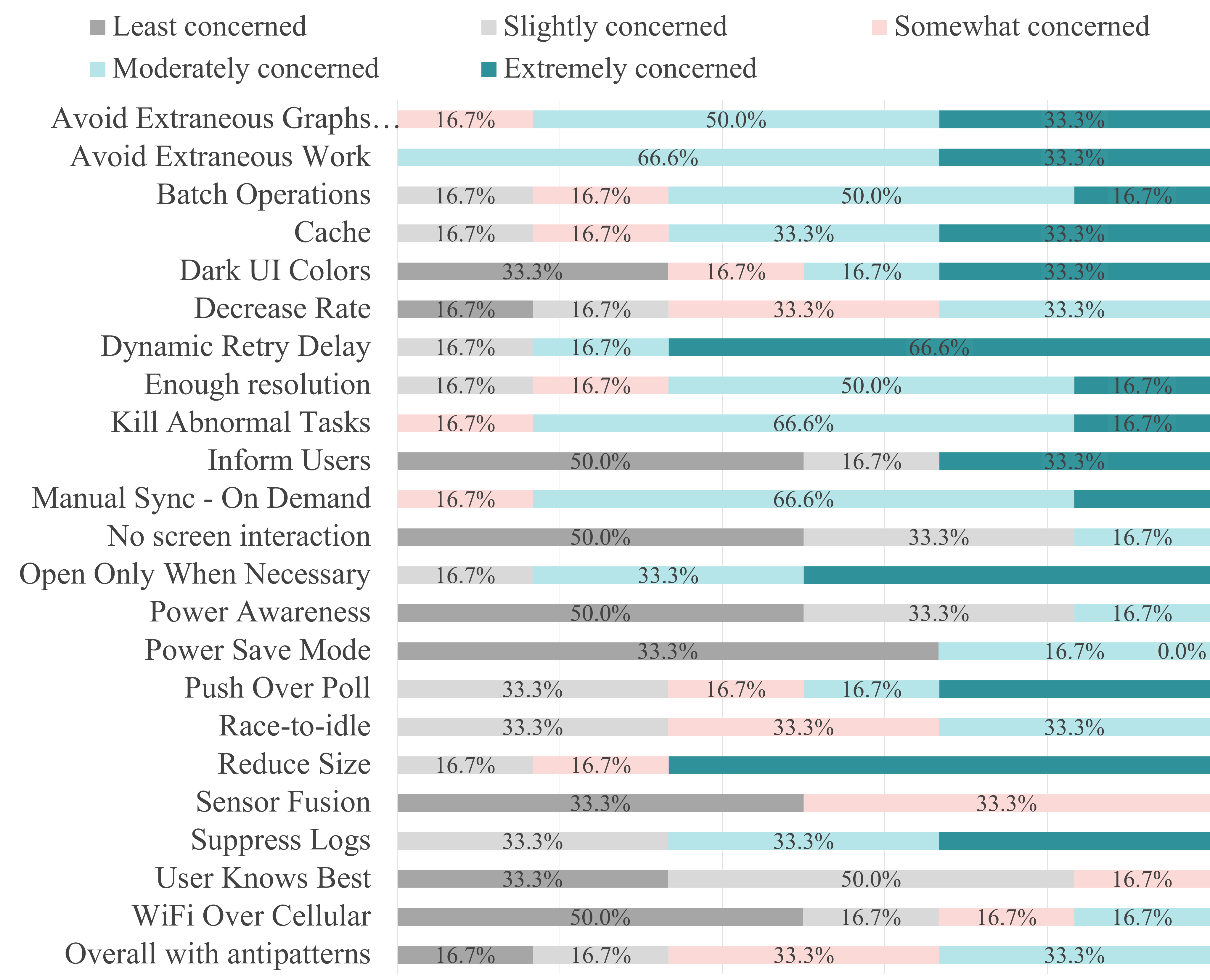}
	\caption{Developers concerns for \eaps}
	\label{fig:rq2-results-concern-overall}	
\end{figure}

\paragraph{Awareness}
\autoref{fig:rq2-results-familarity} shows that most of the interviewed developers
are unfamiliar with \eps. Only a few
possess an understanding of it, primarily gained from university seminars or through their peers. 
Some interviewees highlighted that discussion forums and social media have contributed to their knowledge; this finding suggest that to enhance awareness and influence within the industry, researchers should consider exploring strategies to effectively engage with online social platforms. 

\paragraph{Ported \EPS}
When examining the ported patterns, interviewees found certain patterns easy to understand, such as \singlequote{\darkui}, \singlequote{\pushpoll}, \singlequote{\suplog}, or \singlequote{\drr}.
However, for other patterns, many asked clarification questions and examples as they found them hard to imagine in the Web domain.
This was mainly due to some patterns being applicable to many contexts (\eg \singlequote{\decreaserate}, \singlequote{\reducesize}), hard to pinpoint to an example (\eg \singlequote{\avoidextrawork}), or unheard of (\eg \singlequote{\mansyncod}, \singlequote{\noscrinteraction}). For example, concerning \singlequote{\avoidextrawork}, P5 explained: ``It's hard to pinpoint, this could be anywhere or everywhere.'' For \singlequote{\avoidgraphic}, P3 answered: ``Nothing comes to mind,'' when asked to locate it in the source code. They also asked clarification questions for architecture (\eg ``is it MVC?''), framework, libraries, and programming languages. Although, we restricted our focus on client-server architecture and specific languages (that the interviewees are expert of), such factors seem to play a role in porting them. Furthermore, when discussing patterns, Web developers tend to prioritize application functionality and performance, in contrast to Mobile developers who show greater concern for battery life and cellular network considerations \cite{Luis19a,liu2013has}.

\paragraph{Concerns}
\autoref{fig:rq2-results-concern-overall} reveals significant interviewees are somewhat or moderately concerned with \antipatterns in their code.
 Most developers are especially wary of patterns that could affect the main functionality (functional requirements) of the website, or 
 can have an impact on overall quality or security of software, \eg the \antipatterns for \singlequote{\avoidextrawork}, \singlequote{\dynretdel}, \singlequote{\killtask} \etc  
Many of these patterns recommend avoiding extra work or killing abnormal tasks as they can slow down the performance or even block the main functionality.
On the contrary, interviewees are least or slightly concerned for \antipatterns like  \singlequote{\darkui}, \singlequote{\informusr}, or \singlequote{\poweraware}   
as the patterns do not really impact the main functionality and mainly enhance user experience---having these patterns in source code is not a high-priority tasks for developers.
Some patterns like \singlequote{\raceidle}, \singlequote{\decreaserate} 
seem neutral to developers.

A thematic analysis (excerpts shown in the RP \cite{reppackage})
of developer feedback showed that their concerns (somewhat or moderately concerned with \eaps) stems from the energy demand of cloud computing and the link between computational complexity and \eaps.
Whether such perceptions correspond to reality is currently unknown.
Some developers expressed indifference about energy consumption on local machine; as P5 (who rated `somewhat concerned') put it:
``It depends if its locally run, or on the cloud. If local, I don't really care. If GPUs are involved or the cloud, I am more concerned because I read a lot about energy consumption of large data centers, and I know that GPUs consume a lot of power.''

This feedback indicates the knowledge gap about software energy consumption on the individual level or local machines. 
Given the high number of pet projects built by developers on GitHub \cite{barr2012cohesive}, such systems collectively can have a large impact.
Moreover, some developers prioritize application and performance optimization over energy, as stated by P6 (who rated `somewhat concerned'):
``Application optimization and performance are still more important than energy optimization.''
It calls for more research knowing whether this preference stems from lack of knowledge about energy issues or entrenched focus on application performance throughout the development life cycle.

\input{tables/pattern-guidelines}
\paragraph{Locating Patterns in Code}
Finding these patterns or \antipatterns in source code can be challenging as some are general coding practices and can exist in various components. Developers require a thorough understanding of their web applications.
When inquired about finding these patterns in their code,
\autoref{tab:patterns-guidelines} shows instances of developer guidelines on locating them. Following their guidelines, we indicate (in column \emph{Found}) if a pattern or \antipattern  could be found in the company's source code. For some patterns, they provided a rationale (in \emph{Rationale/Guidelines} column) whether an \antipattern exists in their source code or if it is inapplicable, as shown in the table. 
Not all instances were found, but they could point out potential code components, \eg modules, classes, functions, or specific annotations. Concerning \singlequote{\oown}, one developer (P1) explained:
``Our frontend is a single-page application, It loads everything at once and there is a lot of data traffic to be able to render the page.  We load certain information (in accessing the dashboard) that is not shown to the user (\antipattern), while we don't load dynamic charts (pattern).''

\autoref{fig:rq2-results-resource-invest} reveals developers' interest in optimizing energy aspects.
They expressed willingness to implement \eps in their source; but noted it is often a management decision. They wish to have a dashboard reflecting their source code's energy consumption so that they can identify energy hot spots.
While half of the interviewees stated to be open to refactor existing code, most prefer to incorporate \eps for new source code, \eg in implementing new features or choosing a new library or architecture based on the energy aspect, rather than refactoring existing code for this specific reason.
Indeed, their preference is to address energy concerns spontaneously rather than  allocating specific times, as shown in \autoref{fig:rq2-results-resource-invest}.
When asked about resource commitment, they showed interest in some activities \eg quality assessment, refactoring source code, or identifying \eaps (bug prediction) on a weekly or monthly basis, but  preferred training developers further apart, like quarterly or yearly.

About the knowledge on energy-related tools or past use of such tools , they mentioned a few tools but not intended to explore the energy aspect, \eg \emph{Silk} for data transmission, \emph{tick stack monitoring}, and Profilers for minimizing database requests.
For instance, P6 stated:
``I use the `Network / Performance' tab in Developer Mode in the browser, but it is more for performance or debugging reasons than for energy patterns.''
Such instances show  the limited awareness about existing ones and the dearth of dedicated tools for energy measurement.

\autoref{fig:rq2-results-demographics} shows the demographics of the interviewees in terms of role and experience. More than half of the developers have more than five years of experience in Python and JavaScript while lacking experience in TypeScript.
Also, the majority of developers program frequently (daily or weekly). Only one developer reported once in a year programming activity in TypeScript.  

\subsection{\texorpdfstring{$RQ_3$}{RQ3}: Impact of \EPS} 



 \begin{figure}[ht]
 	\includegraphics[width=\linewidth]{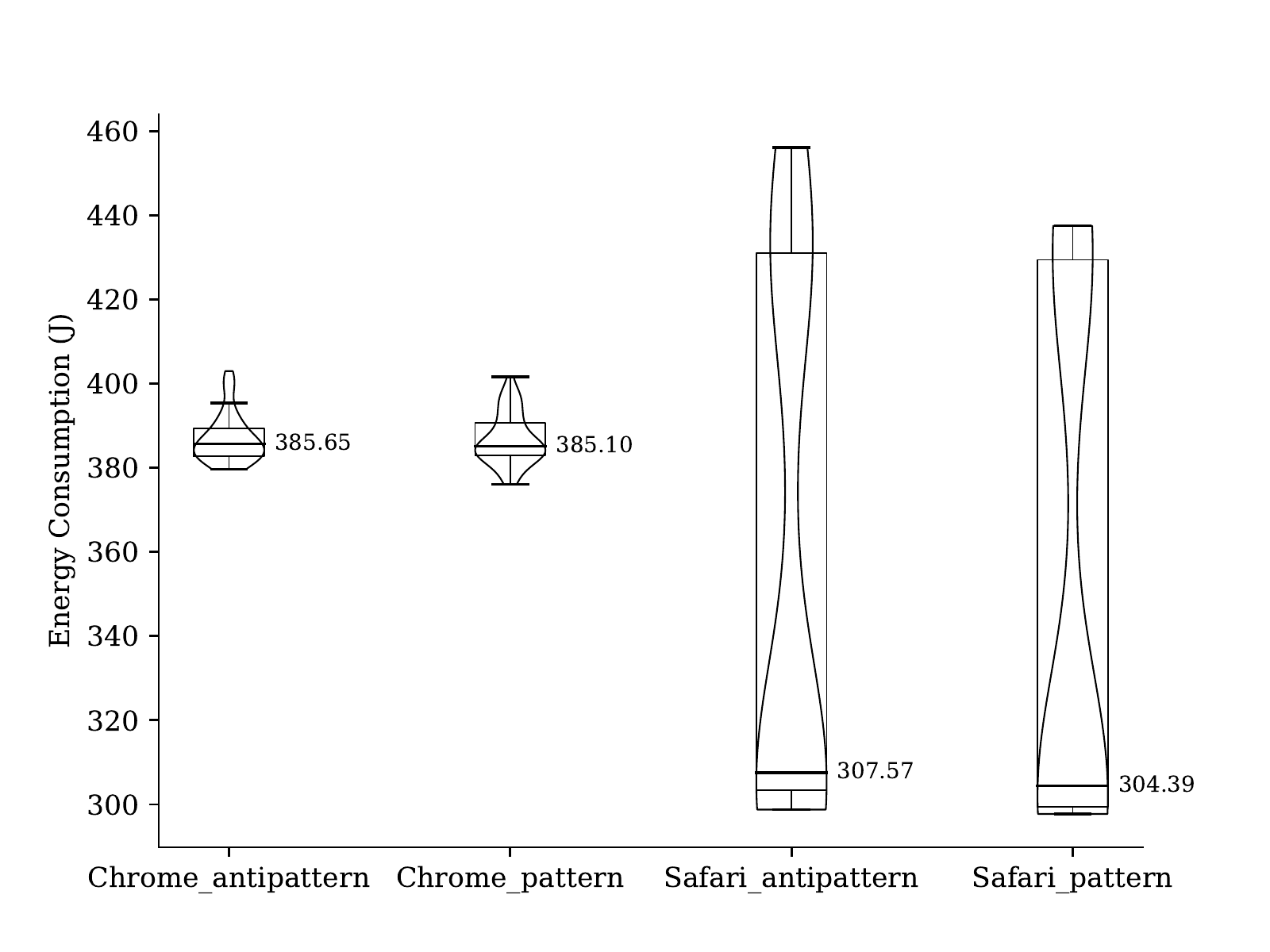}
	\caption{Cumulative energy for \drr}
	\label{fig:rq3-drr-cumulative}	
\end{figure}
 \begin{figure}[ht]
   	\includegraphics[width=\linewidth]{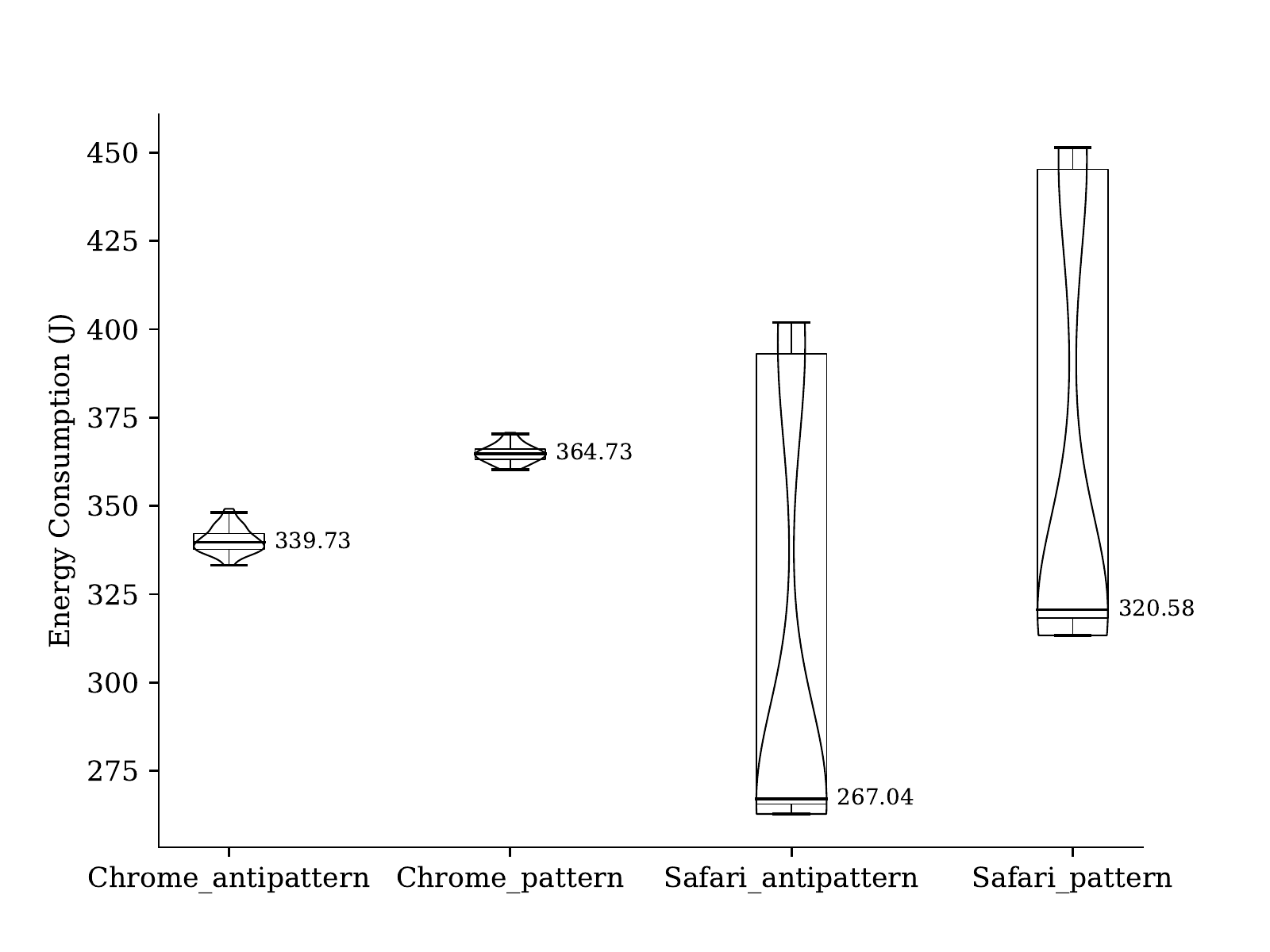}
	\caption{Cumulative energy for \oown}
	\label{fig:rq3-lazyload-cumulative}	
\end{figure}
\autoref{fig:rq3-drr-cumulative} shows the cumulative package energy (CPU) consumption with median values for \emph{\drr} pattern (dynamic interval) and its \antipattern (static interval) for Safari and Chrome browsers.
The energy difference between both scenarios is insignificant, \ie the \antipattern consumes slightly more energy, but we can see a difference between Safari and Chrome browsers. 
Looking at the elapsed time (time to execute the scenario), we found Chrome took an average of approximately 95 seconds, while Safari took approximately 109 seconds. Such differences are mostly due to the Safari WebDriver taking longer to start than the ChromeDriver. However, once started, Safari consumed less energy than Chrome.
We found similar differences in package temperature, where the average temperature for Safari (35 degrees) is less than Chrome's (38 degrees). Previous work has indicated that there is a negative correlation between the high temperature of the device and battery life \cite{ma2018temperature}.

Similarly, \autoref{fig:rq3-lazyload-cumulative} shows the cumulative package energy (CPU) consumption for \singlequote{\oown}. 
We found that the \singlequote{\oown} pattern (lazy loading) consumes more energy than its counter scenario (eager loading).
Also, Chrome (both pattern or its \antipattern) consumes more energy than Safari.
While this may seem counter-intuitive at first glance, since an energy pattern is expected to consume less energy than its \antipattern, the reason can be the implementation of the \singlequote{\oown} scenario. 
To achieve lazy loading of content,  scrolling activities must be constantly monitored to detect when new content needs to be loaded, and several calls to the server must be executed to load the content itself. However, the real advantage of \singlequote{\oown} is that certain resources do not have to be loaded at all if a user does not need them.
In our simulating scenarios, the same number of images (resources) is loaded for both pattern and its \antipattern.

Like \singlequote{\drr}, we observed the elapsed time for \singlequote{\oown} and found that Chrome took an average of approximately 57 seconds, while Safari took approximately 67 seconds. 
We found that similar to \singlequote{\drr}, the average temperature for Safari (43 degrees) is less than Chrome's (45 degrees) for \singlequote{\oown} as well.

Overall, our results show that Safari consumes less energy than Chrome. One of the reasons for this can be that the measurements are done on a machine running macOS, and Safari is optimized for macOS and Apple Hardware. 
Also, Chrome uses V8 JavaScript engine, while Safari uses JavaScriptCore (aka Niro). The way these engines handles JavaScript can affect energy consumption. 
Our results also show that the energy efficiency of \eps may depend on the specific implementation and use case and the frequency of those scenarios. 
Therefore, we advise researchers to carefully design scenarios to measure the patterns.

\section{Implications and Future work}
\paragraph*{For Researchers and Developers} In RQ1, we aim to catalog \eps for Web apps.
We build on the previous work that used \github (Mobile repositories) to identify Mobile \eps
\cite{cruz2019catalog}.
Others have used platforms like Stack Overflow to identify \eps \cite{Shanbhag2022}.
We plan to expand the catalog by exploring other sources like Stack Overflow and web specific \github repositories.
Then, verify the catalog with developers to 
ensure that they are adapted appropriately to the web context.

In RQ3, we measured the energy impact of two patterns.
We observed that adapting patterns requires understanding trade-offs in specific scenario.
For instance, \singlequote{\raceidle} pattern suggests releasing the resources as soon as possible, but holding onto them might save energy in some situations, \eg if the resource request is frequent, then it can consume more energy in the process of requesting and releasing the resources continuously. 
We plan to examine these trade-offs in real-world scenarios. 
Some patterns may save energy at the cost of other software attributes, \eg performance, code comprehension, or maintainability.
Previous work has investigated the effects of code smells, design patterns, or \antipatterns on maintainability \cite{abbes2011empirical}.
They also investigated the effect of energy-efficient changes on maintainability and found that such changes hinder the maintainability of Android applications 
 \cite{Luis19a}. Whether such findings are specific to maintainability or Mobile apps, is yet unknown.
 Knowing such trade-offs can help developers in choosing when to include or avoid patterns in their source code to build green web applications.


\paragraph*{For Tool Designers}  RQ2 findings show developers' interest in exploring the software energy usage, \eg via Dashboard or identify \antipatterns in their source code. 
However, many are unaware of existing tools or find them unsuitable for energy evaluation.
Many developers mentioned tools like \emph{Silk}, or \emph{tick stack monitoring}, 
but use them for debugging or device performance rather than energy analysis.
Future work should explore if this gap is due to unawareness or the need for better tools. 

\paragraph*{For Organizations} The results of RQ2 show that developers learn about energy efficiency from peers and University seminars. 
Developers who are somewhat concerned for \antipatterns are motivated due to company mission or team leaders for achieving sustainability. 
Future work can investigate the awareness and role of the team leaders on source code sustainability.
Companies can leverage this by offering training programs and urging 
team leaders to adopt green coding practices and make other developers in the team aware of them. 
Also, integrating software energy efficiency in the University curriculum can educate students or future developers learn the importance of green software engineering and energy-related trade-offs in development.

\section{Threats to validity}
\paragraph{Catalog of Patterns}
There is a possibility of more patterns that can be suitable to the Web domain. 
We have searched the literature on the existing Mobile patterns and selected the largest catalog by \citet{cruz2019catalog} to adopt.
Also, some patterns are specific to domains and code implementation, while others are more general, as they (\eg \singlequote{\oown}) are relevant across various code components \eg from front-end to database to network components; thus require adaptation to a broader context. We have, for now, kept the patterns generalized and indicated one example for each pattern.

\paragraph{Interview Study} For the interview study, we interviewed three developers from one company and three from different ones. As a result our convenience sample is not representative of web developers. Our choice was dictated by the need to conduct in-depth in-person interviews (60--90 minutes) and to access to their source code to verify their guidelines.
Given that reducing carbon emissions and our energy footprint is a pressing societal need, we cannot rule out that some of the interviewees could have given specific answers due to social desirability bias~\cite{furnham1986response}.
Overall, further studies are needed to both challenge our findings with a more general population (questions to consider include: How familiar are developers with \eps? Do they share similar concerns about \antipatterns? Can one easily locate Web \eps in source code?) and lower the risk of social desirability bias, by means of guaranteed anonymity.
Also our analysis of their feedback can pose a threat. To minimize this issue, two authors independently analyzed the responses and a third author reviewed the disagreements. We handled the disagreements by mutual discussion.

\paragraph{Tools}
As we have conducted the experiment on MacOS, it can introduce a bias on our power consumption estimation. We replicated the experiment additionally on Chrome Browser to minimize this risk. We suggest future work to replicate our work on other operating systems and browsers.
For  \singlequote{\drr} scenarios, we drew inspiration from the EcoAndroid tool and used \powergadget for energy consumption measurement. Any error in these tools can influence our results. 
We tried to minimize this risk by ensuring to be consistent with how these tools have been evaluated and used in previous work.
Although \powergadget is heavily used in previous work, various factors, such as electricity outage, or room temperature can affect energy consumption.
We maintained our conditions consistent, \eg by keeping the desktop computer plugged in and at room temperature for the entire duration of the experiment.
Since the temperature can play a major role in such experiments, we included warm-up and cool-down periods.

\paragraph{User Scenarios}
Having a realistic yet precise simulating scenario is a vital part of energy measurement.
We designed two scenarios per pattern, the scenario of pattern and \antipattern.
To minimize external influences, we removed additional details, \eg in \singlequote{\drr}, API requests contain only a header. Similarly, in \singlequote{\oown} we have selected images as the content to be loaded dynamically. As  \singlequote{\oown} can be applied to other components, \eg database, camera, and thus can have a different energy impact in those scenarios. 
On the Web, there is a constant exchange of network information. PowerLog does not measure network traffic; therefore, we have focused on measuring the more processor-dependent patterns.

\section{Related Work}
 \paragraph*{Energy Patterns}
While previous research in SE has mostly focused on traditional software design issues, such as 
code smells, or design smells \cite{taibi2017developers,bavota2012empirical},  
attention has shifted recently towards energy-related concerns in the source code, leading to identifying \eps in various domains.
For example, Albonico \etal analyzed energy practices in robotics software \cite{albonico2021mining}, finding ten recurring \antipatterns and proposing 14 recommendations to address them. 
Similarly, Shanbhag \etal identified eight \eps for deep learning applications \cite{Shanbhag2022}.
 Although many of their patterns are focused on large language models, patterns such as avoiding unnecessary data referencing is a more generalized recommended practice and similar to ours as \singlequote{\avoidextrawork}, \singlequote{\openwneccesary}, \singlequote{\avoidgraphic}.
 Cruz \etal identified 22 \eps for Mobile-based applications.
 We mapped their \eps to the Web domain, checked them with web developers, and asked developers for guidelines to identify them in source code.  

\paragraph*{Developers' insights on \EPS}
Despite growing interest in the energy aspect, assessing software energy consumption is rare in the industry. 
Previous surveys have shown that developers rarely consider energy efficiency in developing software \cite{Pang2016,Shanbhag2022}. 
\citet{manotas2016empirical} extended the survey to a broad domain of developers from various organizations, such as Microsoft, Google, and IBM.
They found that developers care about energy, but they lack the required information and infrastructure to develop software. 
We confirmed the past research in this regards. We collected their understanding and concerns about \eps.
We inquired for the guidelines to locate these patterns in code.
With the guidelines, we could find eleven patterns and nine \antipatterns in their source code. 

 \paragraph*{Measurement of Software Energy Consumption}
Researchers have analyzed various software or parts of the software to identify energy-hungry practices.
They measured energy consumption of software in various context, such as code execution  \cite{ribic2014energy,pinto2014understanding,carccao2014measuring}, design patterns \cite{maleki2017understanding}, third party libraries \cite{schuler2020characterizing}, dataframe processing libraries \cite{shanbhag2022energy}, programming languages \cite{pereira2017energy}, Java Applications \cite{vijaykrishnan2001energy}, embedded software \cite{tiwari1994power}, Mobile applications \cite{kwon2013reducing}.
As a result, they identified various coding practices or \antipatterns that can cause energy inefficiencies.

Given the increasing usage of Web and ecological cost, its energy efficiency is crucial.
Some organizations have recommended energy-saving practices for Web applications, like optimizing image formats (AVIF, WebP) to reduce file transfer over browsers. 
Greenspector \cite{greenspector} measured the energy impact of these image formats and resolutions for various browsers.
Singh \etal explored the energy cost of Java APIs on servers and found that developers can choose energy-efficient APIs to reduce the energy cost \cite{singh2015impact}.
Inspired by such studies, we measured the impact of two patterns and their \antipatterns in two browsers and found that patterns do not always guarantee less energy consumption and browsers and many other factors play a role.

\section{Conclusion}
Given the increasing usage of web applications, we aimed to define Web \eps to reduce the energy footprint of these applications.
We investigate the porting of existing Mobile \eps to the Web domain and found that 20 patterns could be ported. 
We interviewed developers from different companies to challenge these ported \eps and to investigate their awareness and concerns for energy topics.
While our interviewees were not familiar with energy issues, most expressed concerns for \antipatterns, especially with functional \antipatterns.
Together with their help, we collected guidelines to identify the ported \eps in source code. Based on the guidelines, we could find nine patterns and eleven \antipatterns in the company's source code.
Finally, we provided evidence that, although these patterns aim to save energy, they do not always succeed. 
We measured the impact of two patterns, \singlequote{\oown} and \singlequote{\drr} and found that the \singlequote{\oown} pattern consumes more energy compared to its \antipattern, while we found no evidence for the \singlequote{\drr} pattern compared to its \antipattern.

\begin{acks}
P. Rani and A. Bacchelli acknowledge the support of the Swiss National Science Foundation for the SNF Project 200021\_197227.
\end{acks}


%% file: Listings/avoidExtraWork.tex
\begin{minipage}{0.90\linewidth}
    \begin{lstlisting}[caption={Mozilla API handles visibility change \cite{MozillaAPI}},
     label={AvoidExtraenousWorkCode}]
    const audio = document.querySelector("audio");
    
    // Handle page visibility change:
    // - If the page is hidden, pause the video
    // - If the page is shown, play the video
    document.addEventListener("visibilitychange", () => {
      if (document.hidden) {
        audio.pause();
      } else {
        audio.play();
      }
    });
\end{lstlisting}
\end{minipage}

%% file: tables/patterns-adaptability.tex
\begin{table*}
\centering
\footnotesize
		\caption{Energy patterns with applicability to web, classified to client (C) or Server (S), description, and examples.}
		\label{tab:patterns_adapatability}
		\begin{supertabular}{p{0.16\linewidth}p{0.07\linewidth}p{0.015\linewidth}p{0.68\linewidth}}			
     \textbf{Pattern} & \textbf{Applicability} & \textbf{C/S} &\textbf{Description}  \\ \midrule
Avoid Extraneous Graphics &  \yes & S &  Use battery-intensive graphics or animations with moderation.\\
and Animations & & & \eg A website not loading heavy graphics until users interact with them. \\
\midrule

\avoidextrawork   & \yes      & S           & Present only relevant data or perform tasks that have a direct impact on the user experience. \\
& & & \eg Mozilla's API in \autoref{AvoidExtraenousWorkCode} informs users for the page visibility to let audio/video pause.\\
\midrule

\batchoper   & \yes          & S            & Combine multiple operations to perform batch processing.\\
& & & \eg  Web API from Microsoft to group several operations into a single HTTP request \cite{MicrosoftAPI}.
\\
\midrule

\cache     & \yes            & C                  &  Utilize caching mechanisms to reduce network load.\\
& 
& & \eg A code example to cache an API response in the local storage \cite{reppackage}.
\\
\midrule

\darkui     & \yes           & C            & Provide a web application with the dark UI color theme. \\
& & & \eg Facebook provides an option on the website to switch to a dark theme. \\
\midrule

\decreaserate    & \yes       & S                 & Increase the time interval between requests to the backend.\\
& & & \eg Library website refreshes the book availability only a few times a day. 
\\
\midrule

\dynretdel  & \yes          & S          & Use a systematic retry increasing time interval after each failed attempt to a resource, such as a database, or network. \\
& & & \eg In the Fibonacci series utilize a retry mechanism API to handle abnormal conditions \cite{reppackage}. 
\\
\midrule

\enoresolution    & \yes       & S             & Provide high-accuracy data only when strictly necessary.\\
& & & \eg 
\emph{AVIF} and \emph{WebP} image formats reduces file sizes in browsers \cite{avif,WebP}.\\
\midrule

\informusr      & partially      & C                 &  Inform users of the energy-intensive operations on the website.\\
& & & \eg Autoplay feature on YouTube consumes a significant amount of energy, but the user is not informed.\\
\midrule

\killtask       & \yes      & S        & Provide means of interrupting energy-hungry operations. \\
& & & \eg A timeout to interrupt an abnormal operation \cite{reppackage}.
\\
\midrule

\mansyncod       & \yes    & S      & Perform tasks exclusively when requested by the user.\\ 
& & & \eg YouTube, with Autoplay feature off, plays song only when user clicks on it.\\
\midrule

\noscrinteraction       & partially   & C           & Whenever possible, allow interaction without using the display.\\
& & & \eg Users interacting with music player websites via other methods (\eg audio, keyboard) \cite{reppackage}. \\
\midrule

\openwneccesary   & \yes  & S & Open resources or services only when necessary.\\
 & & & \eg A camera application opens the camera only to capture an image \cite{reppackage}. 
 \\
 \midrule

 \poweraware     & partially     & C                  & Enable or disable certain website functionalities based on the power status.\\
& & & \eg Chrome offers \emph{Energy Saver} mode \cite{ChromeEnergyMode}, which limits background activity and visual effects (\eg animations and videos) when the device's battery reaches 20\%.\\
\midrule

\powersave  & partially & C  & Provide an energy-efficient mode for the website.\\
& & & \eg BooHoo has introduced an energy-saving mode to its website \cite{boohoo}.\\
\midrule

\pushpoll    & \yes & S & Use push notifications to receive updates from resources instead of actively querying resources (i.e., polling).  \\
& & & \eg The server pushing the updates rather than the client requesting them \cite{reppackage}. 
\\
\midrule

\raceidle        & \yes        & S             & Release unnecessary resources and services. \\
& & & \eg Release camera after a video call, unused variables via garbage collection in source code \cite{Garbagecollect}. \\
\midrule

\reducesize  & \yes          & S                  & Transmit only necessary data.\\
& & & \eg  An HTTP request wherein strings bigger than 1KB get compressed \cite{reppackage}. 
 \\
 \midrule

\sensorfusion      & \no        & -              
& Use data from low-power sensors to infer whether new data needs to be collected from high-power sensors.\\
\midrule

\suplog     & \yes           & S           & Keep the logging rates low.\\
& & & \eg The logging level shown in the console output as `warning' \cite{reppackage}.
\\
\midrule

\usrknowbest      & \yes       & C            & Allow users to customize preferences for energy-critical features.\\
& & & \eg BooHoo \cite{boohoo} let users choose between certain features (bright screen and full availability of images). \\
\midrule

\wificell   & \no           & -       
& Delay or disable heavy data connections until the device is connected to a WiFi network. \\
\hline
\end{supertabular}
\end{table*}

%% file: tables/pattern-guidelines.tex
\begin{table*}
\begin{center}
\footnotesize
		\caption{Guidelines/Rationales on finding a pattern (P) or \antipattern (AP), or not applicable (NA) in the company}
		\label{tab:patterns-guidelines}
		\begin{tabular}
{p{0.16\linewidth}p{0.06\linewidth}p{0.53\linewidth}p{0.16\linewidth}} 
  			\textbf{Pattern} &\textbf{Found} & \textbf{Rationale/Guidelines to locate the pattern} & \textbf{Code Reference} \\\midrule

\avoidgraphic 		& P, AP      & P1, P3: ``\emph{we do not use animations, but charts only.''} & Library, component    \\

\avoidextrawork     & P      & P1: ``\emph{we implement custom functions, \eg compute bills which are needed.}'' &  Function  \\

\batchoper          & P      & P1 : ``\emph{we use batch operations in testing, Redis jobs.''} & Functions    \\
& & P2: ``\emph{look for \texttt{promise.all()}''} & Functions \\

\cache              & P      & P3: ``\emph{most API calls are cached, check \texttt{@memorize} annotation.''} * &   Annotation
\\

\darkui				& AP      & P1: ``\emph{it is not implemented.}''. & Module \\

\decreaserate       & NA, P      & P1: ``\emph{we have no sensors so no sync rates (NA).''}    \\
&& P3: ``\emph{there is a refresh token in logging (P).''} & Component \\

\dynretdel          & AP      & P2: ``\emph{there are no delays, only maximum retries for connecting to the API.}'' & Module \\

\enoresolution      & NA, AP       & P1: ``\emph{it is not applicable to us. We use a third-party library that has this feature (NA).''} & Library    \\
& & P2: ``\emph{API responses have high resolution (AP). We have certain resolutions due to legal reasons.''} \\

\informusr          & AP & P1: ``\emph{not implemented currently.''}    \\

\killtask           & P      & P2: ``\emph{we use timeouts to cancel a computation process.''}  & Annotation  \\

\mansyncod          &  P     & P1: ``\emph{we fetch data if user requests.''} & Component  \\

\noscrinteraction	& NA & P1: ``\emph{it does not make sense for our case''}  \\

\openwneccesary     & P, AP      & P1: ``\emph{the frontend is a single page application, we load certain information that is not shown to the user (AP), while we don't load dynamic charts (P).''} & Classes  \\

\poweraware         & NA & P2: ``\emph{it is not applicable to our application.''}    \\

\powersave          & AP & P1: ``\emph{we do not have options for the user.''}    \\

\pushpoll           & AP      & P1: ``\emph{when computation happens in the server, the frontend requests the state of computation constantly.''} &  Component \\

\raceidle           & P      &  P1: ``\emph{we use tear-down functionality in testing (provided by the framework)''} & Module  \\

\reducesize         & AP      & P1: ``\emph{we sometimes send big payloads from the backend to the frontend to deal with.''} & Class  \\


\suplog             & AP      & P1: ``\emph{we do not do enough logging.''} & File  \\

\usrknowbest        & AP      & P1: ``\emph{no options for the user, we show all charts''}  \\


\hline
\end{tabular}
\end{center}
\end{table*}